\documentclass[prd,a4paper,tightenlines,eqsecnum,nofootinbib,preprint,floatfix,showpacs]{revtex4}
\usepackage{epsfig} 
\usepackage{colordvi} 
\usepackage{graphicx} 
\usepackage{bm}

\def\<{\langle} 
\def\>{\rangle}
 
\newcommand{\text}{\rm} 
\def\Tr{{\rm Tr}\,} 
 
\def\Eq#1{Eq.~(\ref{#1})}

\begin{document}  

\vspace*{1.0in} 

\title{A precise calculation of the fundamental string tension\\ 
in $SU(N)$ gauge theories in 
$2+1$ dimensions\vspace*{0.25in}} 

\author{Barak Bringoltz and Michael Teper\\ 
\vspace*{0.25in}} 
 
\affiliation{Rudolf Peierls Centre for Theoretical Physics, 
University  of  Oxford,\\ 
1 Keble Road, Oxford, OX1 3NP, UK \\ 
\vspace*{0.6in}} 

\begin{abstract}
We use lattice techniques to calculate the continuum string 
tensions of $SU(N)$ gauge theories in $2+1$ dimensions. 
We attempt to control all systematic errors at a level that
allows us to perform a precise test of the analytic prediction 
of Karabali, Kim and  Nair. We find that their prediction is
within $3\%$ of our values for all $N$ and that the discrepancy
decreases with increasing $N$. When we extrapolate our
results to $N=\infty$ we find that there remains a discrepancy 
of $\simeq 1\%$, which is a convincing $\sim 6\sigma$ effect.
Thus, while the Karabali-Nair analysis is remarkably accurate
at $N=\infty$, it is not exact.
\end{abstract}
\maketitle 

\section{Introduction}
\label{intro}

Finding an analytic solution to confinement and the mass
spectrum of four dimensional $SU(N)$ Yang-Mills 
gauge theories continues to be a challenge for theoretical 
physics. It is in the large-$N$ limit 
\cite{tHooft}
that the relation between gauge field theories and string 
theory is most natural 
\cite{Polchinski},
and this relation is strengthened by the AdS/CFT 
correspondence. Indeed a solution to the problem
of confinement may shed light on string theories with moderately
strong coupling.

In lower dimensions, gauge theories are of interest for the
same theoretical reasons.
In $D=1+1$ all dynamical degrees of freedom can be removed 
by gauge fixing, and the theory can be analytically solved, with 
linear confinement arising trivially from the Coulomb potential. 
Moving up one dimension to $D=2+1$, makes the theory much more 
complicated. Gluons become dynamical, a linearly
confining potential appears to be dynamically generated,
and the theory appears to be as analytically intractable
as in the four dimensional case. This is unfortunate since 
the $D=2+1$ theory is also phenomenologically interesting: 
through dimensional reduction, it approximates 
the high temperature limit of the four-dimensional theory. 
Indeed, an analytic solution in $D=2+1$ dimensions would be a 
significant step forward, and would perhaps
bring us closer to a solution of the $D=3+1$ case.

While an exact solution to $D=2+1$ gauge theories is not available,
Karabali and Nair have claimed to obtain a very good approximation 
through studying the continuum Hamiltonian, and expressing it in 
terms of color-singlet fields
\cite{KN}.
Truncating the Schr\"odinger equation for their ground state
functional, they obtained the following prediction for the tension 
$\sigma$ of the string that binds distant static sources in the 
fundamental representation:
\begin{equation}
\frac{\sqrt{\sigma}}{g^2N} = \sqrt{\frac{1-1/N^2}{8\pi}}.
\label{KN_sigma}
\end{equation}
Remarkably, this prediction turns out to be within $\sim 3\%$ 
of the lattice calculated values
\cite{Teper_Nd3,Teper_Lucini} 
for all values of $N$.

If one replaces these sources with ones in a general representation, 
${\cal R}$, then the analysis of \cite{KN} predicts that the 
tension $\sigma_{\cal R}$ of the string between these sources, obeys
\begin{equation}
\sigma_{\cal R}=\sigma \, C_{\cal R}, 
\label{casimir}
\end{equation}
where $C_{\cal R}$ is the quadratic Casimir of the representation ${\cal R}$.
In \Eq{casimir} there are no restrictions on  ${\cal R}$, and so this 
clearly cannot be exact in general: in $D=2+1$ gluons screen color
charge and will, for example, break the string if ${\cal R}$ has 
trivial $N$-ality. However in the limit $N\to \infty$  screening
vanishes and, moreover, the lattice calculations
\cite{Teper_Nd3,Teper_Lucini} 
show a decreasing discrepancy with \Eq{KN_sigma} as $N$ increases.
Indeed the extrapolated  $N = \infty$ lattice value is only
about  $\sim 1\%$ below the prediction of \Eq{KN_sigma}.
At this level of accuracy there are a number of systematic 
errors that were inadequately controlled in these lattice 
calculations, and so this leaves open the tantalising possibility
that the calculation of 
\cite{KN}  
might become exact at $N=\infty$. The purpose of the present paper
is to provide a lattice calculation in which all the systematic
errors are controlled at a level that allows us to test this
possibility.

One of these systematic errors has to do with corrections to the
string energy as a function of length. This is an interesting 
theoretical subject of its own, having to do with the nature
of the effective string theory that describes the confining flux 
tube. We expand in detail on this issue in a companion publication 
\cite{NGpap}
and here will only quote a few relevant results. A related issue
is how the string tension varies as a function of the representation 
of the flux. We extend a similar control of systematic errors to
these calculations in a second companion paper
\cite{CSpap}
where we test how well the Casimir scaling in \Eq{casimir}
is satisfied -- in particular by stable $k$-strings.

In the next section we describe the basic lattice setup.
We then describe the methodology used in the calculation of the 
string tension, what are the important systematic errors, and how 
we control them. We then provide our results, extrapolate them to 
the continuum limit, and then extrapolate to $N=\infty$. 
We finish with our conclusions.

Some of our preliminary results, both from this paper and from
\cite{NGpap},
have been presented in 
\cite{lat06pap}.

\section{Lattice setup}
\label{setup}

The theory is defined on a discretized periodic Euclidean three 
dimensional space-time lattice, with spacing $a$ and, typically, 
with $L_s^2L_t$ sites. The 
Euclidean path integral is given by
\begin{equation}
Z=\int DU \exp{\left( -\beta S_{\rm W}\right)},
\label{Z}
\end{equation}
where $\beta$ is the dimensionless lattice coupling, and is related to
the dimensionful coupling $g^2$ by
\begin{equation}
\lim_{a\to 0}\beta=\frac{2N}{ag^2}.
\label{betag}
\end{equation}
In the large--$N$ limit, the 't Hooft coupling $\lambda=g^2N$ 
is kept fixed, and so we must scale $\beta=2N^2/\lambda \propto N^2$
in order to keep the lattice spacing fixed (up to $O(1/N^2)$
corrections). The action we choose to use is the standard Wilson action
\begin{equation}
S_{\rm W}=\sum_P \left[ 1- \frac1N {\rm Re}\Tr{U_P} \right],
\label{eq:SW}
\end{equation}
where $P$ is a lattice plaquette index, and $U_P$ is the plaquette 
variable obtained by multiplying link variables along the circumference 
of a fundamental plaquette. We calculate observables by performing 
Monte-Carlo simulation of \Eq{Z}, in which we use a mixture of 
Kennedy-Pendelton heat bath and over-relaxation steps for
all the $SU(2)$ subgroups of $SU(N)$.

\section{Methodology}
\label{methodology}

To obtain the string tension, $\sigma$, we calculate the energy of 
the lightest flux tube that winds around one of the spatial tori.
Extracting the mass from the correlation function is the
first area in which we need to control systematic errors,
as described below.
 
In the confining phase such a winding flux tube cannot break.
Since there are no sources here (in contrast to Wilson loop 
calculations of the static potential) there are no 
extraneous contributions to the energy (such as the Coulomb
potential) and one can hope to obtain a string description
for all lengths $l$ of the flux tube. There is of course a 
smallest possible length $l_c=1/T_c$, where $T_c$  is the 
deconfining temperature, below which there are no winding flux 
tubes. However for $N\geq 4$ the transition is first order
\cite{LiddleTeper_Tcd3}
and so for larger $N$ we can hope to have a string description
for any $l \geq l_c$. Such a description should become
particularly simple at $N=\infty$ where mixing and decay
vanish. We have performed a careful study of the way the
string energy depends on its length, which will be published 
elsewhere
\cite{NGpap}. 
Here we use those results to bound the theoretical uncertainties 
in extracting the asymptotic string tension from the string energy,
so controlling the associated systematic errors.

There are also systematic errors in extrapolating to the
continuum $a=0$ limit and, subsequently, to the $N=\infty$ 
limit. These will be discussed below as well.

\subsection{Extracting string masses from correlation functions}
\label{excited}

We calculate the energy of the winding flux tube from the 
correlator of suitably smeared $\vec{p}=0$ Polyakov loops 
that wind around a spatial torus. This is a standard 
technique
\cite{Teper_Nd3,Teper_Lucini}
with the smearing/blocking designed to enhance the projection
of our operators onto the ground states. (We use a scheme
that is the obvious dimensional reduction of the one in
\cite{LTW_ops}.)
We calculate with several blocking levels and construct the
full correlation matrix. From this we obtain best estimates
for the ground and excited string states using a variational
method applied to the transfer matrix $\hat{T}=e^{-aH}$ -- again
a standard technique
\cite{var,Teper_Nd3,Teper_Lucini,LTW_ops}.

In practice, our best operator for the string ground state
has an overlap $\sim 99\%$ onto that state so that the
normalised `ground state' correlation function satisfies
\begin{equation}
C(t) = (1-|\epsilon|) \exp\{-M_0(l)t\}
+ |\epsilon_1| \exp\{-M_1(l)t\} + ... 
\quad ; \quad
\sum_{i} |\epsilon_i| = |\epsilon| \sim 0.01
\label{correl}
\end{equation}
where $M_0,M_1$ are the ground and first excited state string
energies. (Since our time-torus is finite, we use cosh fits
rather than simple exponentials, although in practice we
use $L_t$ large enough for any contributions around the
`back' of the torus to be negligible.)
To extract $M_0$ from this correlator one can
fit with a single exponential for $t\geq t_0$, discarding the
lower $t$ values so that a statistically acceptable fit is
obtained. This is a reasonable approach and one followed in  
\cite{Teper_Nd3,Teper_Lucini}.
However it neglects the systematic error arising from the
fact that there is certainly some excited state contribution
as demonstrated, for example, by the fact that one cannot obtain 
a good fit with a single exponential from $t=0$. To control this
systematic error we also perform fits with two exponentials,
with a fixed mass $M^*$ for the excited state, resulting in
a mass $M_0(M^*)$ for the ground state. Typically  $M_0(M^*)$
is smallest when $M^*$ is as small as possible, i.e. $M^* = M_1$,
and is largest when  $M^*=\infty$, i.e. effectively a
single-exponential fit. So the true value typically satisfies:
\begin{equation}
M_0(M_1) \le M^{\rm true}_0 \le M_0(\infty).
\end{equation}
From here on, we refer to the single-cosh fitting procedure by
`S', and to the double-cosh fitting procedure, by `D', and add 
these as superscripts to any
relevant results (such as masses, string tensions, etc.). 
 Consequently we shall have two continuum string tensions
$\sigma^{S}$, and $\sigma^D$, that bracket the true string tension 
\begin{equation}
\sigma^D \le \sigma^{\rm true} \le \sigma^S.
\end{equation}

\subsection{Extracting string tensions from string energies}
\label{sec:Ldep}

From the ground state string energy, $M_0(l)$, we need to extract 
the tension $\sigma$. Taking into account the L\"uscher term 
\cite{old_works},
$\sigma$ is given by
\begin{equation}
\sigma= \frac{M_0(l)}{l} 
+ \frac{\pi}{6l^2}
+  O\left(\frac{1}{l^4}\right).
\label{sigma_def}
\end{equation}
In practice using $\surd\sigma l \geq 3$ one can expect the neglected 
$O(1/l^4)$ corrections to be small. However they represent another
systematic error that needs to be controlled. To do this we shall
use the results of our study in 
\cite{NGpap}, 
where we have calculated  $M_0(l)$ as a function of $l$ for $N=3,4,6,8$.
We have done so in the range 
$1.3-1.6\stackrel{<}{_\sim}l\sqrt{\sigma}\stackrel{<}{_\sim}3-6.2$
and for different lattice spacings. We find that our results can be 
well encompassed by 
\begin{equation}
\left( \frac{M_0(l)}{\sigma l} \right)^2 = 
1  - \frac{\pi}{3\{\sqrt{\sigma} l\}^2} 
- \frac{0.2(1)}{\{\sqrt{\sigma} l\}^5}.
\label{fit}
\end{equation}
Here the first two terms on the right hand side constitute the exact
Nambu-Goto string prediction. It is believed that when one
expands $M_0(l)/{\sigma l}$ in inverse powers of $\surd\sigma l$
the first two corrections are universal
\cite{old_works,new_works}
and equal to those for the Nambu-Goto string. (This is
consistent with our numerical calculations in
\cite{NGpap}.)
The next possible correction term corresponds to the last term 
on the right of \Eq{fit}. We observe that the fitted coefficient is 
much smaller than the $O(1)$ coefficient characteristic of the other 
terms. This shows that Nambu-Goto provides a remarkably accurate 
description of the ground state winding string energy for all  
possible lengths.

The calculations of $M_0(l)$ in this paper are performed for
$l\surd\sigma \geq 3$. We extract the corresponding values 
of $\sigma$ using \Eq{fit}. For such $l$ the contribution of 
the correction terms that are additional to the L\"uscher 
correction is in fact almost negligible, as was assumed in 
earlier calculations
\cite{Teper_Nd3,Teper_Lucini}.
However now we are able to control the accuracy of that
assumption.

\subsection{Extrapolation to the continuum limit}
\label{sec:continuum}

To extrapolate to the continuum limit we need to choose a
theoretically motivated fitting ansatz for the way $a\surd\sigma$ depends 
on the bare lattice coupling $\beta$. From \Eq{betag} it is
clear that 
\begin{equation}
\lim_{a\to 0}\frac{\beta}{2N^2} a \surd\sigma
=
\frac{\surd\sigma}{g^2N}.
\label{betasig}
\end{equation}
The leading perturbative correction to this relation will be 
$O(1/\beta)$. (Note that we do not have here a ratio of physical 
quantities for which the leading correction would be $O(a^2)$.) 
Lattice perturbation theory is notoriously ill-behaved and to reduce
the higher order corrections we replace $\beta$ by the mean-field 
improved coupling 
\cite{parisi_MF}
\begin{equation}
\beta_{\rm MF}=\beta \times \<\mathrm{Tr}U_P\>
\label{betaMF}
\end{equation}
as in
\cite{Teper_Nd3,Teper_Lucini}.

In 
\cite{Teper_Nd3,Teper_Lucini}
the continuum extrapolation was performed with a leading 
$O(1/ \beta_{\rm MF})$ correction. The values of $a\surd\sigma$
from the coarser values of $a$ typically did not lie on the fit,
and were excluded. Although this is a sensible procedure, it
ignores the (small) systematic error due to the neglect of higher 
order corrections in $1/ \beta_{\rm MF}$. Here we will control 
this error by fitting with an additional correction 
\begin{equation}
\frac{a\sqrt{\sigma}}{2N^2}\beta_{\rm
  MF}=\left(\frac{\sqrt{\sigma}}{g^2N} \right)_{\rm continuum} +
\frac{a_1}{\beta_{\rm MF}} + \frac{a_2}{\beta^2_{\rm MF}}. 
\label{continuum}
\end{equation}
By comparing these fits with linear fits where we constrain
$a_2=0$, we shall have an estimate of the effect of higher order 
corrections. Clearly such a strategy is only possible where
the calculations are of sufficient range and accuracy, as in 
the present paper.

It is important to note that this procedure is not without its 
ambiguities. The expansion in \Eq{continuum} is a weak coupling one
which is functionally incorrect in the strong coupling region. 
If our fit includes one or more points in the strong coupling region, 
it is these points that may well determine our estimate of the 
coefficient $a_2$ in \Eq{continuum}, in which case the estimate
will be unrelated to its actual value. This problem is 
exacerbated by the fact that unlike the case in $D=3+1$ (for $N\geq 5$), 
the separation between strong and weak coupling does not involve a
clear-cut first order transition, but rather a cross-over
that turns into a smooth phase transition at $N=\infty$
\cite{fbmt}, 
very much like the Gross-Witten transition in  $D=1+1$
\cite{GW}. 
This crossover peak increases with $N$ and lies in the range 
$\beta/2N^2 \sim 0.40-0.45$
\cite{fbmt}, 
and so in our fits we shall avoid using any values obtained on 
the strong coupling side of this peak.

\subsection{Extrapolation to $N=\infty$}
\label{sec:planar}

The continuum value of $\surd\sigma/g^2N$ is expected to
have a finite limit, with leading corrections of $O(1/N^2)$.
Hence linear fits in $1/N^2$ can be used to extrapolate
to $N=\infty$, as in
\cite{Teper_Nd3,Teper_Lucini}. 
To control the neglected higher order corrections to the linear 
fit we shall also perform fits using the more general form
\begin{equation}
\frac{\sqrt{\sigma}}{g^2N}
=
\left(\frac{\sqrt{\sigma}}{g^2N} \right)_{\rm N=\infty} 
+\frac{b_1}{N^2} + \frac{b_2}{N^4} 
\label{contN}
\end{equation}
and compare the results to those of linear fits ($b_2=0$).
There is no reason to believe that this expansion becomes
functionally incorrect at small $N$, and we shall use it
all the way down to $N=2$.

\section{Results}
\label{sec:results}

\begin{figure}[htb]
\centerline{
\includegraphics[width=14cm]{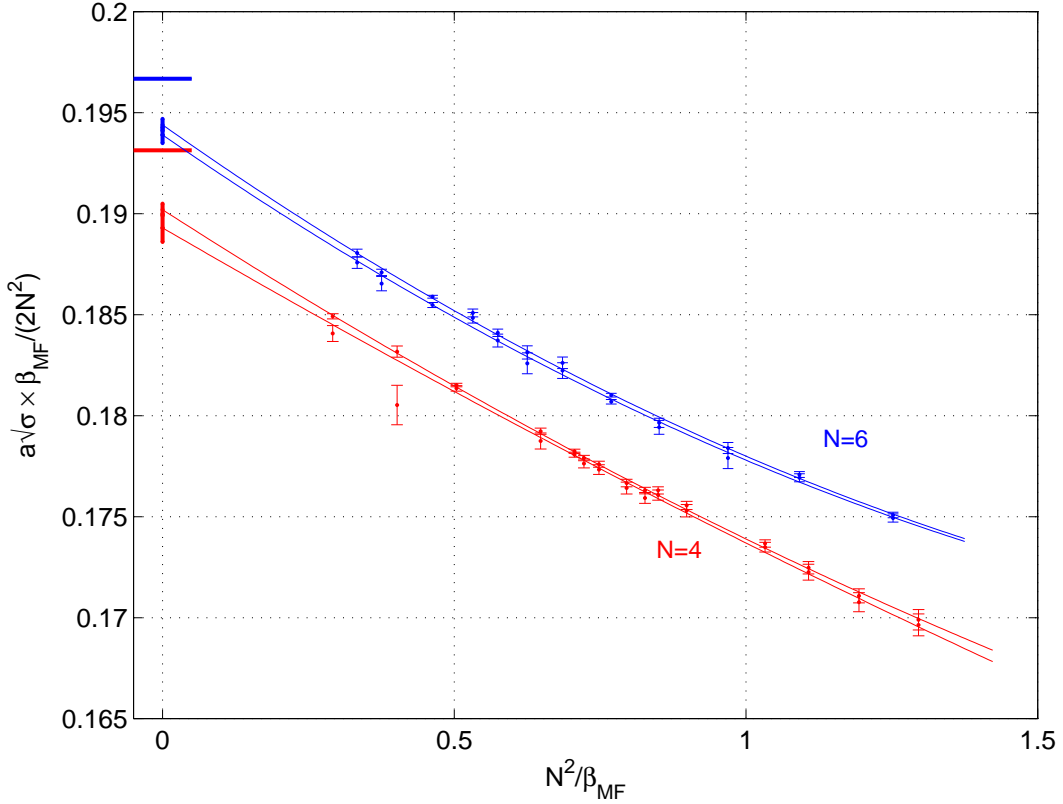} 
}
\caption{The dimensionless quantity 
$\beta_{\rm MF} \frac{a\sqrt{\sigma}}{2N^2} 
\stackrel{a\to 0}{\longrightarrow}  \frac{\sqrt{\sigma}}{g^2N}$ 
as a function of the improved inverse coupling
  coupling $1/\beta_{\rm MF}$ for $N=4,6$. The error bars at
  $1/\beta_{\rm MF}=0$ denote the result of the
  continuum extrapolation, while the horizontal bars denote the values
  predicted by Karabali, Kim, and Nair \cite{KN}.}
\label{fig1}
\end{figure}

In Fig.~\ref{fig1} we plot the values of 
$a\sqrt{\sigma}\beta_{\rm MF}/(2N^2)$, as obtained from both $S$ and 
$D$ fits, versus $N^2/\beta_{\rm MF}$ for our SU(4) 
and SU(6) calculations. We display fits that are 
quadratic, as in \Eq{continuum}. The $O(1/\beta^2)$ correction is
always positive, except possibly for SU(2), so the quadratic
fit leads to a significantly higher value 
than the linear one in the continuum limit.
Despite this fact, we see from the figure that the extrapolated
string tensions are  still lower than the values predicted
by Karabali, Kim, and Nair in \Eq{KN_sigma}.

We give the results of our continuum fits in 
Tables~\ref{table2}-\ref{table3}, where we also list for comparison
the Karabali-Kim-Nair predictions. 

\begin{table}[htb]
\centering{
\begin{tabular}{|c|c|c|c|c|c|c|} \hline \hline
         & $N=2$ & $N=3$ & $N=4$ & $N=5$ &  $N=6$ & $N=8$\\ \hline \hline
$a^S_1$\,& $-0.0133(7)$ & $-0.0158(4)$ & $-0.0186(9)$ & $-0.0175(9)$ & $-0.0204(8)$ & $-0.0186(8)$\\ \hline 
$a^D_1$\,& $-0.0122(12)$ & $-0.0152(9)$ & $-0.0168(19)$ & $-0.0160(14)$ & $-0.0200(11)$ & $-0.0186(10)$\\ \hline \hline
$a^S_2$\,& $-0.0004(3)$ & $+0.0005(2)$ & $+0.0023(6)$ & $+0.0018(6)$ & $+0.0040(5)$ & $+0.0025(5)$\\ \hline 
$a^D_2$,&  $-0.0009(5)$ & $+0.0004(3)$ & $+0.0012(12)$ & $+0.0010(8)$ & $+0.0039(7)$ & $+0.0025(6)$\\ \hline \hline
\end{tabular}
}
\caption{The parameters $a_{1,2}$ in the fit \Eq{continuum}, which are obtained
  for the string tensions $\sigma^{S,D}$. The superscripts $S,D$
  denote the way we fit the correlation function, and bracket the
  actual string tension (see Section~\ref{excited}).}
\label{table2}
\end{table}

\begin{table}[htb]
\centering{
\begin{tabular}{|c|c||c|c|c|c|c|c|} \hline \hline
Type of fit & Source of data & $N=2$ & $N=3$ & $N=4$ & $N=5$ & $N=6$ &
$N=8$\\ \hline \hline
Linear & $M^S$\,& $0.1678(1)$ & $0.1837(2)$ & $0.1892(1)$ & $0.1917(2)$ & $0.1932(2)$ &
$0.1948(2)$ \\ \hline 
Linear & $M^D$\,& $0.1675(3)$ & $0.1831(3)$ & $0.1886(3)$ & $0.1910(4)$ & $0.1927(3)$ &
$0.1943(3)$ \\ \hline \hline
Quadratic & $M^S$\,& $0.1675(3)$ & $0.1839(2)$ & $0.1902(3)$ & $0.1924(3)$ & $0.1944(3)$ &
$0.1955(3)$ \\ \hline 
Quadratic & $M^D$\,& $0.1666(6)$ & $0.1832(4)$ & $0.1893(7)$ & $0.1915(5)$ & $0.1939(4)$ &
$0.1951(4)$ \\ \hline \hline
\multicolumn{2}{|c||}{KKN prediction}& $0.17275$ & $0.18806$ & $0.19314$ & $0.19544$ & $0.19668$ &  $0.19791$ \\
\hline
\end{tabular}
}
\caption{Our continuum values of $\sqrt{\sigma}/g^2N$ and the predictions of
  Karabali, Kim, and Nair (KKN) \cite{KN}. We present results from
  linear and quadratic extrapolations to the continuum ($a_2=0$ or
  $a_2$ as a free fit parameter).}
\label{table3}
\end{table}

\begin{figure}[htb]
\centerline{
\includegraphics[width=14cm]{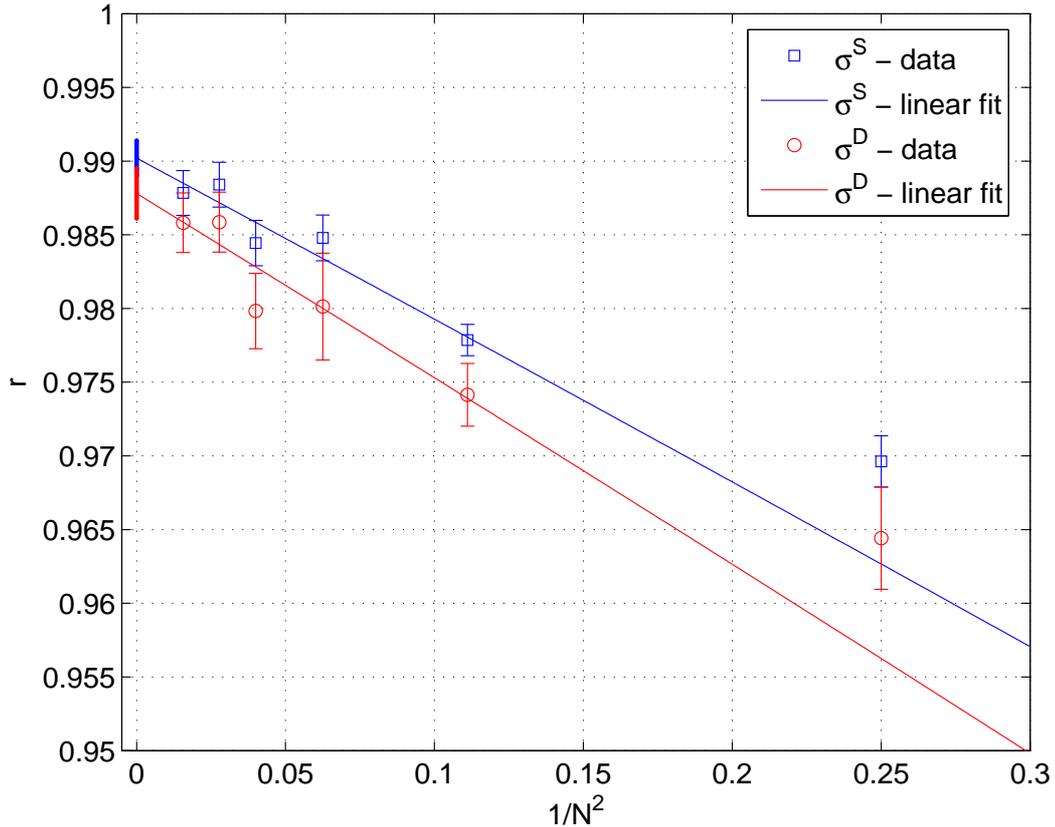} 
}
\caption{The ratio $r$ between the prediction of \Eq{KN_sigma} and 
our data, as a function of $1/N^2$. The error bar at $1/N^2=0$ 
denotes the linear extrapolation to the $N=\infty$ limit.}
\label{fig2}
\end{figure}

Since we are interested in the accuracy of the  Karabai-Kim-Nair 
(KKN) prediction with increasing $N$, we define the ratio $r$
of that prediction to our lattice values
\begin{equation}
r\equiv \frac{\left( \sqrt{\sigma}/g^2N\right)_{\rm
    KKN}}{\left(\sqrt{\sigma}/g^2N\right)_{\rm Lattice}}.
\end{equation}
We now fit $r^2$ with the form
\begin{equation}
r^2=(r_\infty)^2 + \frac{c_1}{N^2} + \frac{c_2}{N^4},
\end{equation}
and present the results for linear fits (with $c_2=0$) and 
quadratic fits (with $c_2$ as a free parameter) in Table~\ref{table4} 
below. (The number of degrees of freedom for all fits was $d.o.f=3$.)
Finally, we show in Fig.~\ref{fig2} the ratio $r$ plotted against
$1/N^2$ with $S$ and $D$ linear fits. As for the continuum
extrapolation, the effect of the higher order correction is to
lift the $N=\infty$ value towards the Karabali-Nair value,
albeit not quite far enough to achieve a perfect match.

\begin{table}[htb]
\centering{ 
\begin{tabular}{|c|c||c|c|c|c|} \hline \hline 
Type of fit & Source of data  & $r_\infty$ & $c_1$ & $c_2$ & $\chi^2/d.o.f.$\\ \hline \hline
Linear & $M^S$  & $0.9902(12)$ & $-0.215(32)$ & - & $0.84$ \\ \hline 
Linear & $M^D$  & $0.9878(17)$ & $-0.245(56)$ & - & $0.64$ \\ \hline \hline
Quadratic & $M^S$ & $0.9908(14)$ & $-0.268(58)$ & $0.41(22)$ & $0.98$ \\ \hline 
Quadratic & $M^D$ & $0.9886(21)$ & $-0.317(103)$ & $0.52(40)$ & $0.57$\\ \hline \hline
\end{tabular}
}
\caption{The extrapolation of $r$ to the large-$N$ limit.}
\label{table4}
\end{table}

\section{Summary}
\label{sec:summary}

We have calculated the tensions of strings in the fundamental 
representation of $SU(N)$ gauge theories in $2+1$ dimensions. 
Our immediate goal was to test the prediction of Karabali-Nair
in \Eq{KN_sigma}, particularly at $N=\infty$, where the screening
effects that are clearly not incorporated in that scheme, vanish.
Since earlier lattice calculations had already shown that any 
discrepancy was no more than a few percent, our
calculation needed to control a number of previously neglected
systematic errors that are small but could be significant at
the $\sim 1\%$ level.

In this paper we described how we controlled the following errors.
Firstly the contribution of excited states to our variationally
selected ground state correlators, from which we extract the
energy of the ground state winding flux loop. Secondly higher 
order string corrections in the relationship between this
ground state energy and the asymptotic string tension.
(Using the results of our companion publication 
\cite{NGpap}
on the effective string theory describing winding flux loops.)
Thirdly, higher order corrections in the continuum extrapolation,
and fourthly, higher order corrections in the extrapolation
in $N$ to $N=\infty$.

Our final results are similar to the ones of the older work
\cite{Teper_Lucini} which assumed that the systematic errors
that we control here, are negligible. We find that this 
assumption is, as it happens, essentially correct. Our string 
tensions are $3\%-1\%$ smaller than the prediction of 
\Eq{KN_sigma}, and a discrepancy persists at $N=\infty$,
where our result is
\begin{equation}
\left(\frac{\sqrt{\sigma}}{g^2N}\right)_{\rm Lattice}
=
0.1975 \pm  0.0002 - 0.0005.
\label{eq:final}
\end{equation}
Here the first error is statistical, and the second 
comes from the difference $\delta \sigma = \sigma_S-\sigma_D>0$.
This error can only lower the string tension, away from
\Eq{KN_sigma}. Consequently, our result is lower by $0.98\%-1.2\%$ 
than \Eq{KN_sigma}
\begin{equation}
\lim_{N\to\infty}\left(\frac{\sqrt{\sigma}}{g^2N}\right)_
{\rm KKN} 
= 
\frac{1}{\sqrt{8\pi}}=0.199471\dots. 
\end{equation}
This difference while small is statistically significant at a 
compelling  $8-5.4$ sigma level (depending on the details of the fit).

While it is clear that the leading term in the scheme of 
\cite{KN}
is not exact at $N=\infty$, our results show that it is astonishingly
accurate.

\section*{Acknowledgements}

MT acknowledges very useful discussions with David Gross
and V.P. Nair during the KITP `QCD and String Theory' Programme
in 2004. BB acknowledges the support of PPARC.
The computations were performed on machines
funded primarily by Oxford and EPSRC.


\begin{thebibliography}{99}
\bibitem{tHooft}
G.~'t Hooft,
  Nucl.\ Phys.\ B {\bf 72}, 461 (1974).
\bibitem{Polchinski}
  J.~Polchinski,
  arXiv:hep-th/9210045.
\bibitem{tHooft2}
  G.~'t Hooft,
  Nucl.\ Phys.\ B {\bf 75}, 461 (1974).
\bibitem{KN}
  D.~Karabali, C.~j.~Kim and V.~P.~Nair,
  Phys.\ Lett.\ B {\bf 434}, 103 (1998)
  [arXiv:hep-th/9804132], and references within.
\bibitem{Teper_Nd3}
  M.~Teper, 
  [arXiv:hep-lat/9804008].
  Phys. Rev. D59:014512, 1999. 
\bibitem{Teper_Lucini} 
  B.~Lucini and M.~Teper,
  Phys.\ Rev.\ D {\bf 66}, 097502 (2002)
  [arXiv:hep-lat/0206027].
\bibitem{NGpap}
B.~Bringoltz and M.~Teper,
in preperation.
\bibitem{CSpap}
B.~Bringoltz and M.~Teper,
in preperation.
\bibitem{lat06pap}
  B.~Bringoltz and M.~Teper,
  arXiv:hep-lat/0610035.

\bibitem{LiddleTeper_Tcd3}
J. Liddle and M. Teper, hep-lat/0509082 and in preparation. \\
K. Holland, hep-lat/0509041.

\bibitem{LTW_ops}
B.~Lucini, M.~Teper and U.~Wenger,
  JHEP {\bf 0406}, 012 (2004)
  [arXiv:hep-lat/0404008].


\bibitem{var}
K.~G.~Wilson, 
Closing remarks at the Abingdon/Rutherford Lattice Meeting,
March 1981,
  K.~Ishikawa, M.~Teper and G.~Schierholz,
  Phys.\ Lett.\ B {\bf 110}, 399 (1982).
  B.~Berg, A.~Billoire and C.~Rebbi,
  Annals Phys.\  {\bf 142}, 185 (1982)
  [Addendum-ibid.\  {\bf 146}, 470 (1983)].
  K.~Ishikawa, M.~Teper and G.~Schierholz,
  Phys.\ Lett.\ B {\bf 110}, 399 (1982).
\bibitem{old_works}
  M.~L\"uscher, K.~Symanzik and P.~Weisz,
  Nucl.\ Phys.\ B {\bf 173}, 365 (1980).
  J.~Polchinski and A.~Strominger,
  Phys.\ Rev.\ Lett.\  {\bf 67}, 1681 (1991).
  J.~F.~Arvis,
  Phys.\ Lett.\ B {\bf 127}, 106 (1983).
\bibitem{new_works}
  M.~L\"uscher and P.~Weisz,
  JHEP {\bf 0407}, 014 (2004)
  [arXiv:hep-th/0406205],
  JHEP {\bf 0207}, 049 (2002)
  [arXiv:hep-lat/0207003].
  J.~M.~Drummond,
  arXiv:hep-th/0608109,
  arXiv:hep-th/0411017.
  N.~D.~Hari Dass and P.~Matlock,
  arXiv:hep-th/0606265.

\bibitem{parisi_MF}
G Parisi in {\it High Energy Physics} - 1980 (AIP 1981).

\bibitem{fbmt}
F. Bursa and M. Teper, hep-th/0511081 to appear in Phys. Rev. D.

\bibitem{GW}
D. Gross and E. Witten, Phys. Rev. D21 (1980) 446.


\end{thebibliography}
\end{document}